\documentclass[aps,prc,twocolumn,groupedaddress,showpacs]{revtex4}
\usepackage{graphicx}
\usepackage[rightcaption]{sidecap}
\begin{document}
\newcommand{\vdot}{\mbox{\boldmath $\cdot$}}
\newcommand{\vect}[1]{\mbox{${\bf #1}$}}
\newcommand{\beq}{\begin{equation}}
\newcommand{\eeq}{\end{equation}}
\newcommand{\etal}{ \emph{et al\/}}
\newcommand{\gvect}[1]{\mbox{\boldmath$ #1$}}
\def\lbar{\mathchar'26\mkern-10mu\lambda}
\newcommand{\hlf}{\mbox{$\frac{1}{2}$}}
\newcommand{\aS}{\mbox{$\arg{S_L}$}\,\,}
\newcommand{\mS}{\mbox{$|S_L|$}\,\,}
\newcommand{\phm}{\phantom{0}}
\newcommand{\ri}{\mbox{${\rm i}$}}
\newcommand{\rri}{\text{i}}
\def\nuc#1#2{\relax\ifmmode{}^{#1}{\protect\text{#2}}\else${}^{#1}$#2\fi}

\title{Evidence for $L$-dependence generated by channel coupling:\\ \nuc{16}{O} scattering from \nuc{12}{C} at 115.9 MeV}
\author{R. S. Mackintosh}
\email{raymond.mackintosh@open.ac.uk}
\affiliation{Department of Physical Sciences, The Open University, Milton Keynes, MK7 6AA, UK}

\date{draft of \today}

\date{draft of \today}
\begin{abstract}
In earlier work, inversion of S-matrix for 330 MeV \nuc{16}{O} on \nuc{12}{C} resulted in highly undulatory potentials; the S-matrix resulted from the inclusion of strong coupling to states of projectile 
and target nuclei. $L$-independent S-matrix equivalent potentials for other explicitly $L$-dependent potentials have been found to be undulatory. Here we investigate the possible implications of the undulatory DPP for an underlying $L$-dependence of the  \nuc{16}{O} on \nuc{12}{C}  optical potential.  We employ S-matrix to potential, $S_{L} \rightarrow V(r) $, inversion yielding local
potentials that reproduce the elastic channel S-matrix of coupled channel (CC) calculations, here applied
to the S-matrix for 115.9 MeV \nuc{16}{O} on \nuc{12}{C}.
In addition, $S_L$ for explicitly $L$-dependent potentials are inverted and the resulting $L$-independent potentials are compared with the undulatory potentials found for \nuc{16}{O} on \nuc{12}{C}.
In this way, we can simulate certain undulatory features of the potentials modified by channel coupling for 115.9 
MeV \nuc{16}{O} on \nuc{12}{C}.

\end{abstract}

\maketitle
\par
\section{INTRODUCTION}\label{intro}
The possibility that $L$-dependence might be a generic property of the nucleus-nucleus optical model potential, OMP, is unwelcome.  It would obviously be inconvenient since standard direct reaction codes would require modification.  This would not be straightforward since there is a great multiplicity of ways in which the OMP might be $L$-dependent. 

Nevertheless, the possibility that $L$-dependence is a general property of the optical model potential 
should be considered. Arguments for the $L$-dependence  of  OMPs for nucleons and some light ions 
have been presented  in Ref.~\cite{arxivL}. Here we raise the question of $L$-dependence for heavier ions, specifically for the case of \nuc{16}{O} on \nuc{12}{C} at a laboratory energy of 115.9 MeV. The discussion 
is based on the result of applying $S_L \rightarrow V(r)$ inversion to the elastic scattering S-matrix $S_L$ calculated by Ohkubo and Hirabayashi~\cite{OH2} in the course of explaining remarkable features in the elastic scattering angular distribution.

In Section~\ref{defs} we briefly present some definitions relating to the IP method~\cite{MK82,ip2,kukmac,arxiv,spedia} for $S_L \rightarrow V(r)$ inversion that will be useful for discussing the results of the calculations.

Section~\ref{thepot} presents and discusses the undulatory potentials found by inverting
$S_L$ of Ohkubo and Hirabayashi~\cite{OH2} for \nuc{16}{O} on \nuc{12}{C} at a laboratory energy of 115.9 MeV. 

Motivated by the results of Section~\ref{thepot}, Section~\ref{model} presents and discusses the $L$-independent potentials found by inverting $S_L$ produced by potentials having a specific model $L$-dependency. This affords the opportunity to compare certain feature of the potentials found in this way with those presented
in Section~\ref{thepot}. The model calculations involve the same nuclei  and the same energy as those of Ref.~\cite{OH2}.
The key comparison is between: (i) the potentials found by inverting $S_L$  calculated with channel coupling and,
(ii) the $L$-independent potentials that have the same $S_L$ as potentials having a known $L$ dependency.

Section~\ref{disco} discusses the results, and also makes some comments as to why it appears to be 
possible to avoid the issue of $L$-dependence in many cases of elastic scattering.

Section~\ref{summa} briefly summarizes the findings.  Throughout this text, the partial wave angular momentum of spinless projectiles will be denoted by upper case $L$.

\section{INVERSION CODE IMAGO: DEFINITIONS}\label{defs}
We present here some definitions that will be used when we discuss results from the 
$S_L \rightarrow V(r)$ inversion code Imago~\cite{imago}:
\begin{description}
\item[IP, SRP] The Iterative-Perturbative, IP, inversion method~\cite{MK82,ip2,kukmac,arxiv,spedia} starts 
the iterative inversion process with a starting reference potential, SRP.  
\item[IB, SVD] At each iteration, amplitudes for
the elements of the inversion basis, IB, are determined using Singular Value Decomposition, SVD,
matrix operations.
\item[S-matrix distance, $\sigma$] After a sequence of iterations, the current potential can be plotted by Imago and compared with the SRP. The fits to the S-matrix $S_L$ and to the angular distribution can also 
be plotted.  Imago will associate different lines on the graphs with values of the `S-matrix distance' $\sigma$ which is defined below in Eqn~\ref{sig}.
\item[Target S-matrix] The `target S-matrix'  is the input S-matrix that is to be inverted. 
\end{description}

The quantity $\sigma$ is defined in terms of two sets of S-matrix elements (SMEs),
the $S^1_L$ and $S^2_L$ as follows:
\begin{equation}
\sigma^2 = \sum_L |S^1_L -S^2_L|^2 \label{sig}
\end{equation}
In most cases, $S^1_L$ will be the the target of the inversion and $S^2_L$ will be the
the S-matrix for the current stage of the inversion. Successful inversions often result in
values of $\sigma$ that are three orders of magnitude lower than that for $S_L$ calculated
from the SRP. (We remark that all of the above can be generalized to spin-half inversion of $S_{lj}$ determining an interaction with a spin-orbit term. Spin-1 inversion leading to a tensor interaction is also possible.) In order to achieve
very low $\sigma$, Imago allows the lower limit on the singular values of the SVD linear system to be progressively
lowered; initial high values of this limit are required to avoid divergence. For further discussion of $S_L \rightarrow V(r)$ inversion see Ref.~\cite{spedia}.

Standard practice when inverting $S_L$ with the code Imago is to compare results with different choices of IB, SRP and other parameters in order to establish the uniqueness of the potential. In some cases, as discussed below, it becomes difficult to identify a unique potential..

\section{THE POTENTIAL FOUND BY INVERSION}\label{thepot}
\begin{figure} [tbh]
\includegraphics[keepaspectratio,width=9.0cm,angle=-90] {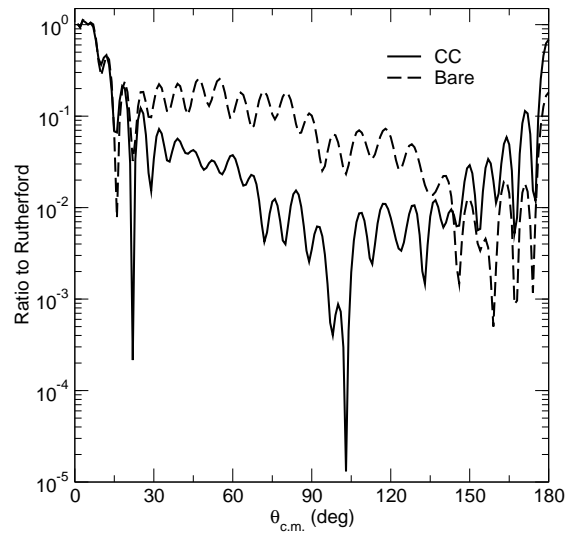}
 \protect\caption{\label{fig1} {
Elastic scattering angular distributions for 116 MeV \nuc{16}{O}  on  \nuc{12}{C}. The dashed line is for  scattering
from the bare potential with no coupling. The solid line is for the full coupled channel calculation of Ref.~\cite{OH2}. } }
\end{figure}

\begin{figure} [tbh]
\includegraphics[keepaspectratio,width=10.0cm,angle=-90] {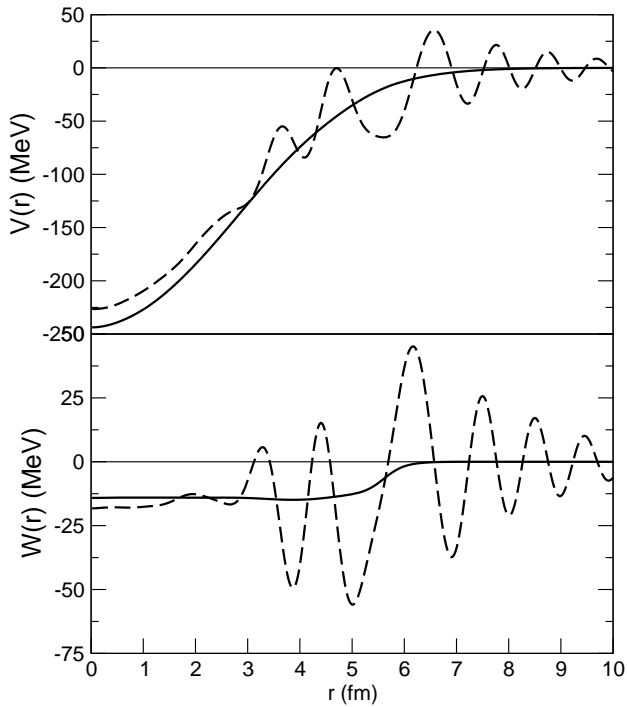}
 \protect\caption{\label{fig2} {
The inverted potential CC2potx fitting $S_L$ for 115.9 MeV \nuc{16}{O} scattering from $^{12}$C, the real part
is in the upper panel and the imaginary part in the lower panel.
The solid line  is for the SRP which is the bare potential.  The dashed line is for the inverted potential 
CC2potx with inversion $\sigma = 4.34 \times 10^{-4}$. } }
\end{figure}

The 115.9 MeV case is very interesting for several reasons, just one of them being
the remarkable meteorological connections~\cite{OH2}. The coupling to states of both projectile and target nucleus specified in Ref.~\cite{OH2} greatly modifies the elastic scattering angular distribution, AD, see Fig.~\ref{fig1}.
The coupling leads to $S_L$ for the elastic scattering channel that, when inverted, yield strongly undulatory (`wavy' or `oscillatory')  potentials. The first result, potential `CC2potx', shown in Fig.~\ref{fig2} corresponding to a low value of S-matrix distance  $\sigma = 0.43 \times 10^{-3}$. For such low values of
 $\sigma$, the AD calculated with the inverted potential  would be indistinguishable from the coupled channel, CC,  AD in a figure such as Fig.~\ref{fig1}, for all angles. We note that for the SRP, $\sigma =0.875$ so that the inversion process has reduced $\sigma$
by more than three orders of magnitude. The SRP in this and most cases is the `bare' elastic channel potential, 
and its $S_L$
corresponds to the  channel coupling being switched off.  Therefore, the difference between the solid and dashed
lines is a representation of the dynamic polarization potential, DPP, that is due to the coupling.

There are three obvious questions: 1. What do these very strong undulatory features mean? 2. Are they realistically a possible property of a nucleus-nucleus single channel interaction potential? 3. Is the potential CC2potx a unique solutions to the inversion problem?

\begin{figure} [tbh]
\includegraphics[keepaspectratio,width=10.0cm,angle=-90] {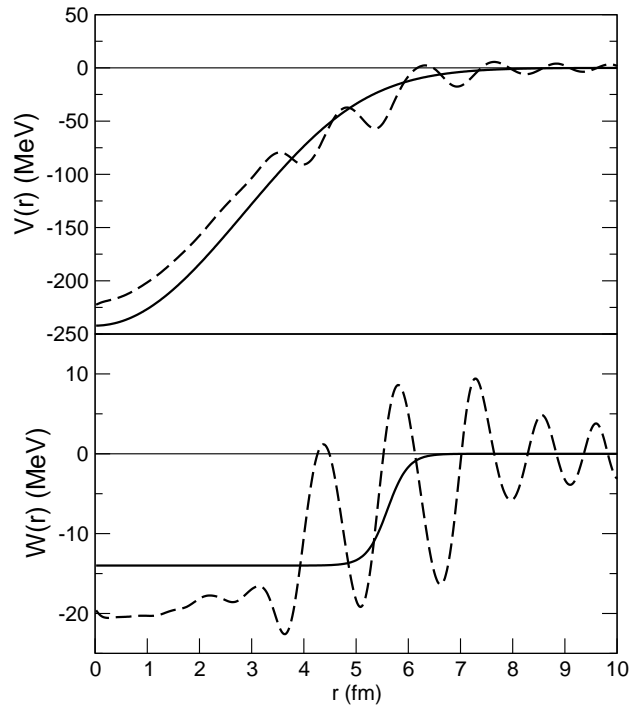}
 \protect\caption{\label{fig3} {The inverted potential CC3pot7 fitting $S_L$ for 115.9 MeV \nuc{16}{O} scattering from $^{12}$C,  the real part
is in the upper panel and the imaginary part is in the lower panel.
The solid line  is for the SRP, the dashed line is for the inverted potential CC3pot7,  $\sigma = 1.38 \times 10^{-4}$. 
   } }
\end{figure}

\begin{figure} [tbh]
\includegraphics[keepaspectratio,width=10.0cm,angle=-90] {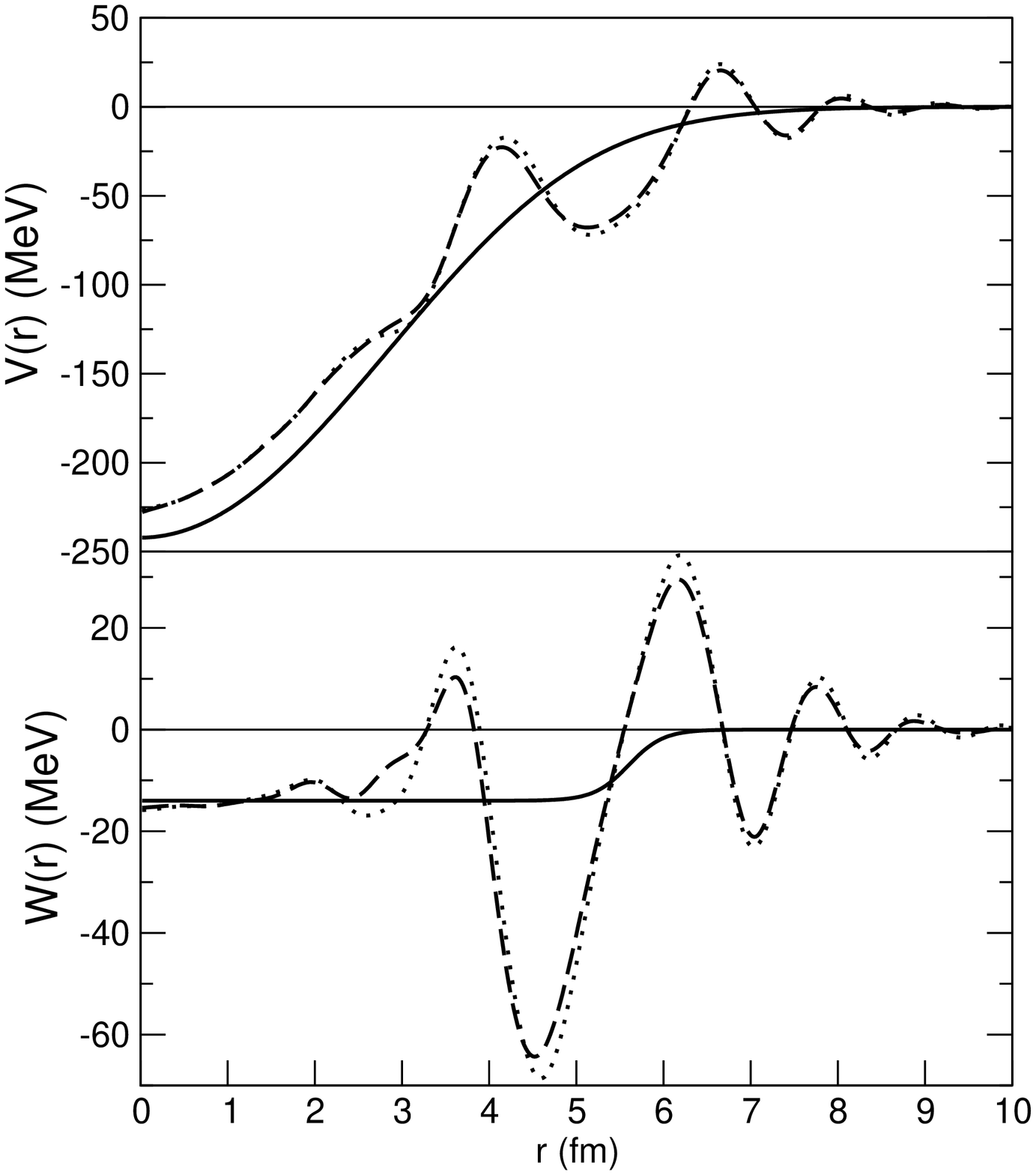}
 \protect\caption{\label{fig4} {The inverted potential CC4pot7 fitting $S_L$ for 115.9 MeV \nuc{16}{O} scattering from $^{12}$C, the real part
is in the upper panel and the imaginary part is in the lower panel.
The solid line  is for the SRP (the bare potential), the dotted line is for the inverted potential CC4pot7,  $\sigma = 1.01 \times 10^{-3}$, and the dashed line is  the potential for an earlier iteration,  $\sigma = 1.55 \times 10^{-3}$. 
  } }
\end{figure}

\begin{figure} [tbh]
\includegraphics[keepaspectratio,width=9.0cm,angle=-90] {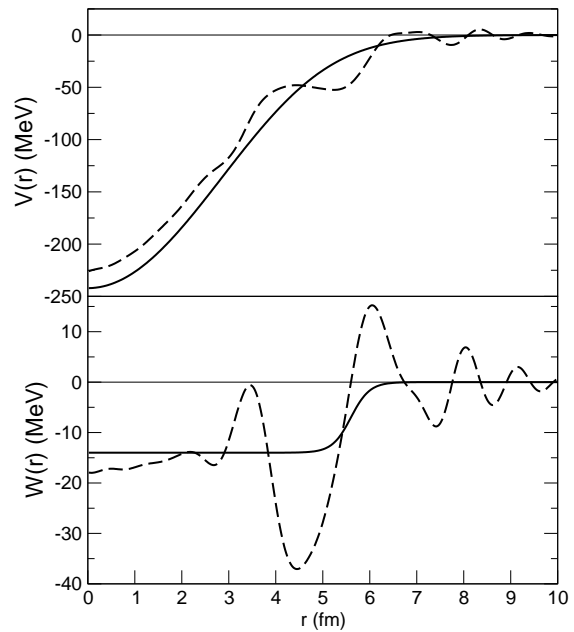}
 \protect\caption{\label{fig5} {The inverted potential CCXpot12 fitting $S_L$ for 115.9 MeV \nuc{16}{O} scattering from $^{12}$C, the real part
is in the upper panel and the imaginary part is in the lower panel.. The solid line  is the SRP (the bare potential)  and  the dashed line is the inverted potential,  $\sigma = 1.03\times 10^{-3}$.  Note that the vertical scale for the imaginary 
part is different from that in Fig.~\ref{fig4}.} }
\end{figure}

The answer to the third question is `no', it is not a unique solutions to the inversion problem as we shall see from the existence of alternative solutions in Figures~\ref{fig3}, ~\ref{fig4} and~\ref{fig5}. In cases where the S-matrix corresponds to a reasonably smooth potential, IP inversion yields a practically unique solution. This  can be established by comparing solutions with different SRPs, different IBs (IBs using different dimensionalities and sets of basis functions) and other parameters. This is usually straightforward, and spurious oscillations can be eliminated.  However spurious oscillations are possible because of the existence of `transparent potentials'. A transparent potential is an oscillatory potential that, when added to an existing potential, leads to very small (effectively zero) changes in the S-matrix and hence the observables. These can be eliminated from IP inversion, except when the true potential that is sought is also highly oscillatory, 
in which case  there is no natural `smoothest' potential. That seems to be the case here. The problem is considerably less severe  for \nuc{16}{O} on \nuc{12}{C} at higher energies, as in the 330 MeV case~\cite{OH1,3author} where there is a larger number of partial waves to determine the potential. Moreover, coupling effects tend to become somewhat weaker at higher energies.

The answer to the second question is that strong undulations are indeed a property of a nucleus-nucleus interaction that includes the effect of strong inelastic couplings as  in the present case.

Concerning the four solutions presented here:  the imaginary part of CC2potx (Fig.~\ref{fig2}) has extreme undulations which extend to a radius far beyond 10 fm. Potential CC3pot7 (Fig.~\ref{fig3}) has much less extreme undulations in the imaginary part, but they also extend unrealistically far out. It was found that solutions which do not extend to unrealistically large radii can be found as in CC4pot7 (Fig.~\ref{fig4}, the dotted line), 
but apparently at the cost of large excursions in the imaginary part and a higher value of 
$\sigma = 1.01 \times 10^{-3}$, about $7\times$ higher than for CC3pot7. In general, following any 
sequence of iterations, the undulations increase as $\sigma$ falls, and this can be seen in Fig.~\ref{fig4} where the dashed line represents the potential for an earlier iteration, with  $\sigma = 1.55 \times 10^{-3}$. The tendency for
the undularity to increase as $\sigma$ falls is evident. An independent inversion (involving an alternative initial inversion basis) led to the potential CCXpot12, shown as the dashed line in Fig.~\ref{fig5}, which has the same overall shape as potential in Fig.~\ref{fig4} and comparable inversion $\sigma$. The AD corresponding to the CCXpot12 is  graphically indistinguishable from the AD for the CC calculation. For CCXpot12, the inversion $\sigma = 1.03 \times 10^{-3}$. The undulations of the potential in Fig.~\ref{fig5}, while having the same general shape as those in Fig.~\ref{fig4}, are of much smaller amplitude, noting the different scale for the imaginary term. The real components
of the potentials in Fig.~\ref{fig4} and Fig.~\ref{fig5} have almost the same volume integral in spite of the different
amplitude of the undulations, and both have a similar increase in rms radius compared to the bare potential. In fact,
all of the potentials of Figures~\ref{fig2} ---~\ref{fig5} exhibit the same uniform repulsive effect for a radius of less than  about 3 of 4 fm. For the imaginary components, the volume integrals and rms radii  of the potentials in Fig.~\ref{fig4} and 
Fig.~\ref{fig5} are quite similar, and in both cases the volume integrals are greater and the rms radii are less than those of the bare potential. We consider CCXpot12 in  Fig.~\ref{fig5} to provide the provisional best potential representation of the elastic scattering S-matrix from the CC calculations~\cite{OH2}. All the potentials presented in Fig.~\ref{fig2} through Fig.~\ref{fig5} yield ADs  that are graphically indistinguishable  out to $180^{\rm o}$ from the CC ADs on the scale of the figures.

Less undulatory potentials inevitably have higher inversion $\sigma$ and fit a limited angular range. For example,
a potential on the iterative path to CCXpot12 with $\sigma$ a factor of 10 times larger had similar general features:
the imaginary term had a dip at 4 fm and a peak at 6 fm although somewhat less pronounced.  The angular distribution for this potential had clear differences from that for the CC calculation beyond $\sim 80^{\rm o}$. These differences were typically a factor of 3 for many angles and a factor of 6 at $180^{\rm o}$.

It is now clear that we have not  provided a unique answer to the following basic question: what $L$-independent potential corresponds to the S-matrix generated by the relevant channel coupling?
However, it is quite certain that no non-undulatory, $L$ independent potential could ever give anything 
approaching a good fit to the angular distribution that is calculated from the S-matrix generated by
the coupled channel model of Ref.~\cite{OH2}. It is very reasonable to assume that the same
would apply to the experimental data that the calculations of Ref.~\cite{OH2} approximately fit.
We therefore give a tentative answer to question 1 above. The strong undulations present an alternative: 
either the potential for the case in question is indeed highly undulatory, or it is $L$-dependent (it might be both). 
The question 
of how generic this alternative is must be the subject of further work; the question of why smooth and 
$L$-independent potentials are so often considered acceptable will be commented upon in Section~\ref{disco}.

In the course of determining potentials that reproduce the CC S-matrix, fits of a precision are required (and achieved)
that are not approached in conventional phenomenology.  However, there is a possible relevance to conventional phenomenology:  the ambiguities, corresponding to transparent potentials, that have been found for 
115.9 MeV \nuc{16}{O} scattering on \nuc{12}{C}, are likely to be present in model-independent fits (spline, sum of Gaussian etc) to precision experimental AD data of wide angular range.

\section{MODEL L DEPENDENCE}\label{model}
In Ref.~\cite{arxivL} some examples were given, for light ion cases, of the relationship between $L$-dependent potentials 
and the corresponding $L$-independent potentials with the same $S_L$
or, in the case of nucleons the same $S_{lj}$. We now present a preliminary exploration of the relationship 
between $L$-dependence and the corresponding undulatory nature of $L$-independent S-matrix-equivalent 
potentials. The examples all relate to the scattering of
\nuc{16}{O} from \nuc{12}{C} at a laboratory energy of 115.9 MeV so as to maximize the relevance
to the situation in Section~\ref{thepot}. 

The idea is to take $L$-independent potentials, impose $L$ dependence upon them and invert
the resulting $S$ matrix in order to study undulations that might arise in those $L$-independent
potentials that are equivalent, in terms of $S_L$, to the $L$-dependent potentials. The $L$-independent potentials 
that we start with have simple Woods-Saxon forms. The real part is chosen to be roughly like 
the bare folded potential of Ref.~\cite{OH2} and the imaginary part is exactly
the imaginary Woods-Saxon (WS) term given in that reference.  The WS parameters for the real part are
$V= 250$ MeV, $R= 3.0$ fm and $a=0.65$ fm and for the imaginary part $V=14$ MeV, $R=5.6$ fm and $a=0.3$ fm.

The imposed $L$-dependence is simple and takes the form of added terms $v(r) \times f(L)$  or $w(r) \times f(L)$ 
where the $f(L)$ factor multiplying a real ($v(r)$) or  imaginary  ($w(r)$) terms is given by:
\begin{equation}
f(L) = \frac{1}{1 + \exp{((L^2 -{\cal L }^2)/ \Delta^2)}} \label{lfac}.
\end{equation}
In the present calculations, $v(r)$ and $w(r)$ each have a Woods-Saxon form with  the same radius and diffusivity parameters as the corresponding real and imaginary $L$-independent terms. As a result,  the $L$-dependent potentials essentially have a renormalized real or imaginary component for $L$ less than $\cal{L}$, with a fairly sharp transition 
since $\Delta$ is quite small. The potential is unmodified for values of $L$ substantially greater than $\cal{L}$. 
This pattern is  motivated by the tendency for undulatory potentials to arise particularly when there is a substantial
change for partial waves having $L$-values around the point where $|S_L| \sim \hlf$. 

\subsection{Including an $L$-dependent imaginary part}
We adopt an $L$-dependent factor $f(L)$ with ${\cal L} =20$ and $\Delta=2$ and $w(r)$ with a  depth of 0.7 MeV. The effect is to increase the depth of the imaginary part from 14 MeV to 14.7 MeV, a 5 \% increase, for values
of $L$ less than 20, with a fairly sharp transition  to zero change for higher $L$. For $L={\cal L}$, $|S_L|\sim 0.35$.

\begin{figure} [tbh]
\includegraphics[keepaspectratio,width=9.0cm,angle=-90] {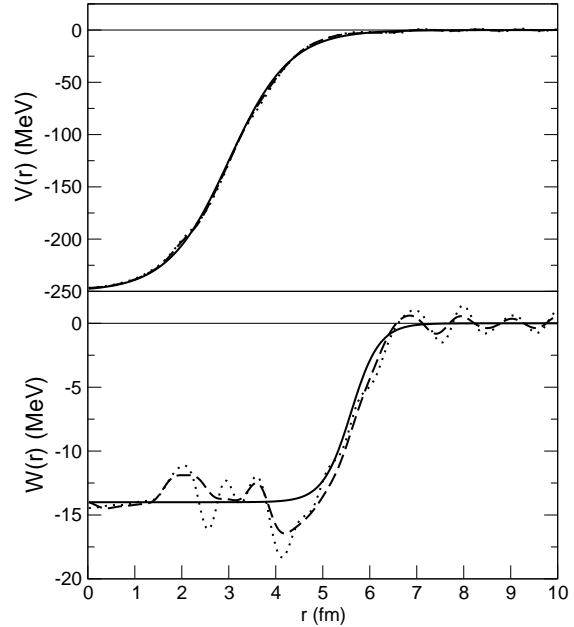}
 \protect\caption{\label{fig6} {The solid lines presents the real  (upper panel) and imaginary (lower panel)  parts of the
$L$-independent potential. The dashed and dotted lines present inverted potentials, reproducing $S_L$ calculated with an imaginary $L$-dependent term defined in the text. Two  inverted potentials are presented, the dotted line with  $\sigma = 0.378 \times 10^{-3}$ and the dashed line, an earlier iteration with  $\sigma = 0.523 \times 10^{-3}$.
   } }
\end{figure}

The inversion results of are shown in Fig.~\ref{fig6} where  the SRP for the inversion is given by the solid line and this is the $L$-independent potential.  In this figure the dotted line corresponds to inversion $\sigma = 0.378 \times
10^{-3}$ and the dashed line for an earlier iteration  corresponds to inversion $\sigma = 0.523 \times
10^{-3}$. It will be seen that the potential has a characteristic oscillatory feature in the surface which has an amplitude much greater than the potential there. This is quite similar to the surface feature in the imaginary potential in Fig.~\ref{fig4}.  The volume integral of the imaginary term is, in each case,  greater than that of the $L$-independent potential,
by 8.56 \% for the dashed case and 8.24 \% for the better fitting dotted case. This is nearly twice the increase imposed on the potential for values of $L$ less than 20. The change in the volume integral of the real part is 100 times smaller, 
amounting to a  $\sim 0.2$\% change. In this sense  the $L$ dependence in the imaginary component has induced a relatively small change in the real component. However, although the imposed undulations are small on the scale of the figure, the magnitude is  comparable to that of the undulations in the imaginary component, but they have a nearly zero volume integral.

The two potentials shown  follow  the general tendency for the undulations to become enhanced in amplitude as $\sigma$ becomes smaller as the iterations progress, i.e.\ as $S_L$ for the $L$-independent potential determined by inversion  more closely approaches $S_L$ for the $L$-dependent potential. Local regions of emissivity in the surface undulations, do not lead to a breaking of the unitarity limit since the well-fitted $S_L$ are calculated to have $|S_L| < 1$. We conclude that such excursions into local emissivity are not an argument against the potentials that we showed in Fig.~\ref{fig4} which exhibit similar undulations in the surface.   We remark that local regions of emissivity are a very common feature of DPPs representing coupling effects in
a wide range of nuclear scattering cases. They also occur in some model independent fits to elastic scattering~\cite{arxivL}.

Naturally, there is no reason to expect that the model $L$-dependence we have applied is a realistic representation of the effects of strong channel coupling. However, it does show that $L$-dependence appears to be
necessary to reproduce the effects of strong coupling with a smooth potential. Probably quite a small
degree of $L$-dependence is enough.

\subsection{Including an $L$-dependent real part}
Similar calculations were performed including an $L$-dependent real part for the same scattering case
and with the same $L$-independent term. Again, the parameter ${\cal L}$ was 20. 
\begin{figure} [tbh]
\includegraphics[keepaspectratio,width=9.0cm,angle=-90] {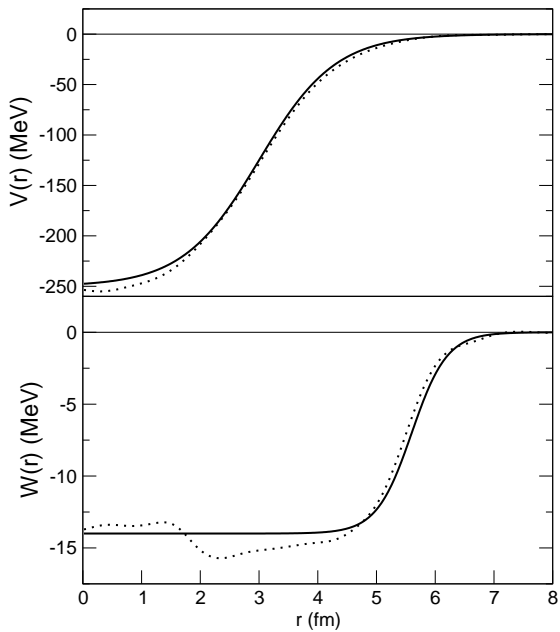}
 \protect\caption{\label{fig7} {Inversion of $S_L$ from the enhanced (see text) $L$-dependent real term. The solid lines represent the real part (upper panel)  and imaginary part (lower panel) of the SRP which is the unmodified the $L$-independent potential. The dotted line is the inverted potential with  $\sigma = 0.799 \times 10^{-3}$.    } }
\end{figure}

The real potential was first increased from 250 MeV to 251 MeV for $L$ less than 20, with the same fairly sharp transition. This represents a small percentage change in the potential for low $L$, although in absolute magnitude it was comparable to the change in the imaginary part, (note however that the radial extent of the real potential is rather less than that of the imaginary term.) The inverted potential had no strong surface undulations, unlike the case with the imaginary
$L$-dependence. The real potential had two regions where the potential is increased in depth: near the origin and near the surface. The real volume integral is increased by 0.59 \%, which is not unreasonable in view of the fact that the potential was increased by 0.4 \% for the lowest 20 partial waves, which are those with substantial penetration. The volume integral of the imaginary potential fell by just 0.19 \%, the positive and negative excursions  
roughly cancelling. 

To obtain a more visual result, $v(r)$ was increased by a factor of 10 so that the real potential was 260 MeV deep  
for $L$  less than $\sim20$  with the same fairly sharp transition. We therefore expect
roughly $10 \times$ the effect. The $L$-independent potential shown as a dotted line in  Fig.~\ref{fig7},  does indeed 
depart from the solid line following the same pattern enhanced in magnitude by that factor. The increase in depth of 
the real potential in the lowest radial range and also around $r = 5$ fm is clear. The change in the 
volume integral is a 5.61 \% increase. The change in the imaginary part is qualitatively like that for the 1 MeV case,
but much larger point-by-point. However the change in volume integral is very small indicating that the positive and negative changes cancel in the integration for $J_{\rm I}$. It appears to be a general rule that while a real $L$-dependent term  leads to a perturbed real $L$-independent term having a substantial change in volume integral, the change in the imaginary term has a small volume integral although not small point by point. 
The converse is also true respecting an imaginary $L$-dependent term. 

Because the unmodified real potential had a much smaller radial extent than the imaginary potential, that
difference applied also to the added  $L$-dependent term; this might relate to the absence of surface 
undulations in  Fig.~\ref{fig7}. 

\subsection{Choice of $L$-dependency}
There are too many possible forms of $L$ dependence for an exhaustive study here, but we have  shown that $S_L$ from the CC calculations of Ref.~\cite{OH2}  implies  $L$ dependence in
both the real and imaginary potentials generated by the coupling. Other forms of $L$-dependency have been applied to heavy ion interactions, and those forms that do not involve a distinct change change between high and low partial wave, as in Eq.~\ref{lfac} when ${\cal L} \sim L_{\rm t}$
with $|S_{L_{\rm t}}|\sim 0.5$, may not lead to strong undulations.

An example of $L$-dependence in the real part is provided by the RGM calculations of Wada and
Horiuchi~\cite{WH} for \nuc{16}{O} + \nuc{16}{O} elastic scattering. The $L$-dependence arises 
from exchange terms beyond the 1-particle knock-on exchange that is normally included implicitly in
folding models. Horiuchi~\cite{horiuchi} reviews such calculations in the context of a more
general discussion of microscopic nucleus-nucleus potentials.  The set of
$S_L$ values corresponding to the $L$-dependent real potentials of Wada and Horiuchi have been
inverted~\cite{ATMCW} to yield an $L$-independent potential which is significantly different at
lower energies from that derived~\cite{WH} using WKB methods. The difference between the
equivalent complete $L$-independent potential from the $L$-independent (non-exchange) part of the Ref.~\cite{WH} potential is most marked in the nuclear interior. This work clearly
established that exchange processes lead to an $L$-dependence of nucleus-nucleus interactions
(in addition to any parity-dependence.) The $L$ dependence  of Wada and Horiuchi apples to partial waves that would,
in a more realistic calculation, be strongly absorbed. This makes their $L$ dependence difficult to establish or disprove 
experimentally.

The model for \nuc{16}{O} + \nuc{16}{O} scattering of Kondo \emph{et al}~\cite{KRS}, included
a phenomenological $L$-dependent real term inspired by the model of Wada and Horiuchi, together
with an $L$-dependent imaginary term. The $S_L$ for the potential with both terms $L$-dependent was readily inverted~\cite{ATCM} and the resulting real potential had a very similar shape and energy dependence to that found~\cite{ATMCW} for the Wada-Horiuchi potential. 

The $L$-dependence of the real part of the Kondo \emph{et al}~\cite{KRS} potential was of the an
overall factor $V_0 + V_1 L(L+1)$, i.e. a gradual $L$ dependence unlike that in Eq.~\ref{lfac}. 
This, by design, leads to a very similar energy dependence for the $L$-independent potential found
by inverting the  Wada and Horiuchi~\cite{WH} S-matrix. It seems that there is a systematic
qualitative difference between the equivalent $L$-independent potentials found for these `gradual'
$L$-dependencies and the sharper Fermi-form $L$-dependencies which tend to generate much more undulatory equivalent potentials.

We remark that a consistent calculation involving both the full antisymmetrization and full channel coupling would be very difficult. However,  it does seem that the effects of antisymmetrization affect the potential
at a small radius and might therefore be less relevant when the strong absorption of realistic calculations
makes effects near the nuclear centre less significant.

\section{DISCUSSION}\label{disco}
The coupled channel calculations of Ref.~\cite{OH2} have three important characteristics: (i) they were based 
on an established cluster model for the interacting nuclei, (ii) they explained an otherwise unexplained  
large angle feature of elastic scattering, and (iii) an $L$-independent potential that reproduces their angular distribution in a single channel calculation has strongly undulatory features. these features include emissive regions, particularly in the nuclear surface. Such effects are likely to elude approximate fits to experimental AD data that is of limited angular range.

The calculations presented here lead to the following conclusion that applies at least to the scattering 
case of our example: in order to generate, \emph{using a potential without undulations},  the S-matrix  $S_L$ that reproduces the effect of channel coupling,  the potential must be $L$-dependent. In other words, in some cases at least, the representation through a potential model of the effects of strong channel coupling 
presents a choice between a potential that is undulatory and one that is $L$-dependent. The same alternative
has previously been firmly established for nucleons scattering from \nuc{4}{He} and \nuc{16}{O} and
other cases~\cite{spedia}. For cases such as that considered here, the range of possible $L$ dependencies makes
it difficult to pin down the specific form of the $L$ dependence of the potential. 

In the case of nucleon scattering, it has been shown that coupling to collective states of the target nucleus generates a DPP having substantial undulations~\cite{mk90}. Formal theory shows~\cite{feshbach,satchler,rawit87,NM14} that the DPP is both $L$-dependent and non-local in a complicated way that is unlike exchange non-locality. However there are separate lines of evidence for both $L$-dependence and the appearance in empirical potentials of departures from the smooth forms  of customary parameterized or folding model potentials~\cite{arxivL,rapaport}. For an example of an 
empirical deuteron potential showing wavy features, see Ref.~\cite{ermer}. There is a need for more such model
independent good fits to elastic scattering data that are both precise and of wide angular range. The concept of a `good fit' is highly context dependent, but in the present context, `good fit' means $\chi^2/{\rm DF} \sim 1.$

The question arises: why has $L$-dependence not been widely accepted? For the case of proton scattering, it is only 
when precise fits to wide angular range data are demanded that the need for either waviness or $L$-dependence 
becomes evident. It is also  the case that for nucleon elastic scattering  from target nuclei away from closed shells, angular distributions tend to be smoother  and, as a result, have less power to discriminate between potentials. For the case of heavier ions, it becomes difficult to measure angular distributions over a wide angular range. The angular distribution fitted in Ref.~\cite{OH2} for  \nuc{16}{O} on \nuc{12}{C} at a laboratory energy of 115.9 MeV extended out to about 140$^{\circ}$. This angular range would obviously be impossible for identical bosons. For the case of  \nuc{16}{O} on \nuc{16}{O} at 350 MeV, experimental uncertainties appear to become large at around 55$^{\circ}$. It has to be conceded that potentials with no hint of undulations can fit this angular distribution with an imaginary term that is very close to that which corresponds to the static model Glauber potential~\cite{CM94}; it would be very interesting to know what would be found if more precise data of greater angular range were fitted. Indeed, it would seem \emph{a priori} that the contribution found in \nuc{16}{O} on \nuc{12}{C} scattering at 115.9 MeV~\cite{OH2} due to the excitation of states of \nuc{16}{O} would also contribute to \nuc{16}{O} on \nuc{16}{O} scattering at a similar energy. 

The present calculation raises, but throws little light on, an important general question: how does the supposed $L$-dependence depend upon energy? There certainly were undulatory features in the DPPs that were found at 330 MeV~\cite{3author}, but less severe, and the DPPs were more uniquely determined. For proton scattering, the $L$ dependence appears to fall with increasing energy~\cite{KM79} and this may well be a general property.

\section{SUMMARY}\label{summa}
The $L$-independent potentials that yield the same $S_L$ as the strong channel coupling 
in the case of 115.9 MeV \nuc{16}{O} on \nuc{12}{C} elastic scattering, Ref.~\cite{OH2},  exhibit strong undulations. 
This is firmly established although the exact nature of these undulations is hard to pin down definitively.
An oscillatory potential  in the surface region appears to be required and local excursions into emissivity do not
necessarily lead to the breaking of the unitarity limit.. 
Simple model $L$-dependent potentials lead to elastic scattering S-matrices $S_L$ that, when inverted, 
yield undulatory  $L$-independent  potentials.  The undulations exhibit some of the same features that 
are characteristic of the potentials generated  by collective coupling, including the  oscillations  in the surface.   
There is therefore  a non-exclusive alternative: the potential is undulatory or it is $L$ dependent. 
We conjecture that this would be found to apply generally if serious model-independent fitting of suitable
elastic scattering data were carried out systematically.

There are very many ways in which $L$-dependence could be introduced into a phenomenological
potential, and in Section~\ref{model} we have restricted the choice  to a single form each for the real and
imaginary terms separately. It was found that an $L$-dependent real part generated some moderate 
waviness in both real and imaginary terms of the corresponding $L$-independent potential, but the 
volume integral of the imaginary part was almost unchanged. Conversely, an $L$-dependent imaginary
term left the volume integral of the real part almost unchanged, although that was perturbed point by
point. The $L$-dependence in the imaginary part generated very wide amplitude undulations in the 
corresponding  $L$-independent term. The underlying $L$-independent 
potential, upon which the $L$-dependency was based, had an imaginary term that extended much
further in radius, and it might be that this had an influence of the lesser tendency for $L$-dependence
in the real part to generate large amplitude undulations. This is just one of the many things that could 
be explored in a similar way.

\section{ACNOWLEDGEMENTS}
I am deeply indebted to Professors S. Ohkubo and Y. Hirabayashi for supplying S-matrix elements and
other numerical data, and also for very insightful comments. I am also very grateful to Nicholas Keeley for
producing publishable figures, and also for many helpful discussions.

\end{document}